\documentclass[%
 aip,
 amsmath,amssymb,
 reprint,%
]{revtex4-2}

\newcommand\beq{\begin{equation}}
\newcommand\eeq{\end{equation}}

\usepackage{graphicx}
\usepackage{dcolumn}
\usepackage{bm}

\usepackage[utf8]{inputenc}
\usepackage[T1]{fontenc}
\usepackage{mathptmx}
\usepackage{etoolbox}

\makeatletter
\def\@email#1#2{%
 \endgroup
 \patchcmd{\titleblock@produce}
  {\frontmatter@RRAPformat}
  {\frontmatter@RRAPformat{\produce@RRAP{*#1\href{mailto:#2}{#2}}}\frontmatter@RRAPformat}
  {}{}
}%
\makeatother

\begin{document}

\preprint{AIP/123-QED}

\title[]{Synthesizing quasi-bound states in the continuum in epsilon-near-zero layered materials}
\author{Giuseppe Castaldi}%
\author{Massimo Moccia}
\author{Vincenzo Galdi}
 \email{vgaldi@unisannio.it}
\affiliation{ 
Fields \& Waves Lab, Department of Engineering, University of Sannio, I-82100, Benevento, Italy
}%

\date{\today}


\begin{abstract}
Bound states in the continuum (BIC) are highly confined, nonradiative modes that can exist in {\em open} structures, despite  their potential compatibility and coupling with the radiation spectrum, and may give rise to resonances with arbitrarily large lifetimes. Here, we study this phenomenon in layered materials featuring {\em epsilon-near-zero} constituents. Specifically, we outline a systematic procedure to synthesize quasi-BIC resonances at given frequency, incidence angle and polarization, and investigate the role of certain critical parameters in establishing the quality factor of the resonances. Moreover, we also provide an insightful phenomenological interpretation in terms of the recently introduced concept of ``photonic doping'', and study the effects of the unavoidable material loss and dispersion. Our results indicate the possibility to synthesize sharp resonances, for both transversely magnetic and electric polarizations, which are of potential interest for a variety of nanophotonics scenarios, including light trapping, optical sensing and thermal radiation.
\end{abstract}

\maketitle

In 1939, von Neumann and Wigner\cite{vonNeumann:1993um} put forward a celebrated thought experiment featuring a quantum-mechanical potential that could counterintuitively support bound electronic states with energy residing above the continuum threshold. Such states, typically referred to as ``bound states in the continuum'' (BIC) or ``embedded eigenvalues'', have never been experimentally demonstrated in quantum mechanics, but have recently gained a growing attention in other wave-based disciplines, as a route to synthesizing highly localized resonant modes with large quality-factors ($Q$) in open structures. Specifically, in optics and photonics, BIC-type phenomena have been demonstrated, theoretically and experimentally,  in a variety of configurations involving periodic structures (such as gratings and photonic crystals), waveguides, metamaterials and metasurfaces; the reader is referred to Refs. \onlinecite{Marinica:2008bs,Plotnik:2011eo,Hsu:2013oo,Silveirinha:2014tl,Monticone:2014ep,lannebere:2015om,Kodigala:2017la,Rybin:2017hq,
	Bezus:2018bs,Koshelev:2018am,Monticone:2018tl,Sakotic:2020be} for a sparse sampling, and to Refs. \onlinecite{Hsu:2016bs,Krasnok:2019ai,Azzam:2021pb} for comprehensive reviews.

A series of interesting approaches\cite{Silveirinha:2014tl,Monticone:2014ep,Monticone:2018tl,Sakotic:2020be}
to attain BIC-type phenomena rely on media characterized by ``epsilon near zero'' (ENZ) behavior, i.e., vanishingly small relative dielectric permittivity. Such behavior is exhibited by several natural and artificial materials, and is known to strongly enhance the light-matter interactions in a variety of scenarios, ranging from nonlinear effects to quantum physics (see, e.g.,  Refs. \onlinecite{Liberal:2017nz,Wu:2021en} for a review). In particular, a recent study by Sakotic {\em et al.} \cite{Sakotic:2020be}  demonstrated a class of BIC-type high-$Q$ resonances in ENZ layered materials, which can exhibit strong absorption and thermal emission in narrow spectral/angular regions.  

Here, with focus on symmetric ENZ tri-layers, we revisit the above results, and derive a systematic synthesis procedure for BIC-type high-$Q$ resonances, at a given frequency and angle of interest, for both transversely magnetic (TM) and electric (TE) polarizations. Moreover, we carry out a series of parametric studies to illustrate the effects of certain relevant parameters. For the TM polarization, we also provide an interesting phenomenological interpretation based on the recently introduced concept of  ``photonic doping''.\cite{Liberal:2017pd} Finally, we explore the effects of realistic (lossy, dispersive) implementations.

%
\begin{figure}
	\centering
	\includegraphics[width=\linewidth]{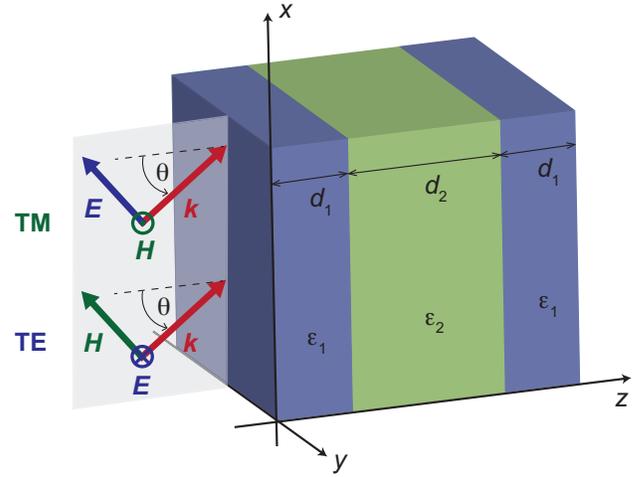}
	\caption{Problem geometry (details in the text).}
	\label{Figure1}
\end{figure}

Referring to the two-dimensional (2-D) geometry in Fig. \ref{Figure1}, with all fields and quantities independent of $y$, we consider a symmetric structure made of three layers, stacked along the $z$-direction and of infinite extent in the $x-y$ plane. Specifically, we assume the structure embedded in vacuum, and identify a {\em cladding} region (the two exterior layers of thickness $d_1$ and relative permittivity $\varepsilon_1$)  and a {\em core} region (the inner layer of thickness $d_2$ and relative permittivity $\varepsilon_2$). We assume that the materials are non-magnetic (i.e., relative permeability $\mu=1$) and, for now, we neglect material dispersion and losses. Moreover, we consider a time-harmonic plane-wave illumination, with suppressed $\exp\left(-i\omega t\right)$ time dependence, angle of incidence $\theta$, and TM ($y$-directed magnetic field) or TE ($y$-directed electric field) polarization. Accordingly, the relevant components of the incident wavevector can be expressed as
\beq
k_x=k\sin\theta,\quad
k_z=k\cos\theta,
\eeq
where $k=\omega/c=2\pi/\lambda$ is the vacuum wavenumber, with $c$ and $\lambda$ denoting the corresponding wave velocity and wavelength, respectively. In what follows, without {\em a priori} assumptions on the ENZ character of the cladding/core regions, we analytically derive the exact conditions for the emergence of a BIC, i.e., the coalescence of a zero and a pole in the optical response (reflection/transmission coefficient).\cite{Krasnok:2019ai}  It is worth highlighting that, in principle, BIC-type phenomena can also be observed in structures as simple as bi-layers (see, e.g., Ref. \onlinecite{Savoia:2014to}), but (at a given frequency) only for incidence from one side, since the reflection responses for incidence from the left ($z<0$) and right ($z>0$) are different. Conversely, in the (symmetric) three-layered configuration considered here, it is possible to attain a BIC-type response for incidence from either side.  

We start considering the TM polarization, which is know to exhibit strong interactions with ENZ layers, mediated by very narrow polariton resonances.\cite{Alu:2007en}  For the structure in Fig. \ref{Figure1}, via some lengthy but straightforward algebra, the reflection coefficient (for incidence from either side) can be expressed as
\begin{widetext}
	\beq
	{R_{^{TM}}} = \frac{{2{\varepsilon _1}{\varepsilon _2}{k_{z1}}{k_{z2}}{t_1}\left( {k_{z1}^2 - \varepsilon _1^2k_{z}^2} \right) + {t_2}\varepsilon _1^2k_{z1}^2\left( {k_{z2}^2 - \varepsilon _2^2k_{z}^2} \right) + {t_2}t_1^2\left( {\varepsilon _1^4k_{z}^2k_{z2}^2 - \varepsilon _2^2k_{z1}^4} \right)}}{{{t_2}{a_1}{a_2} - 2{\varepsilon _1}{\varepsilon _2}{k_{z1}}{k_{z2}}\left( {{\varepsilon _1}{k_{z}}{t_1} + i{k_{z1}}} \right)\left( {{\varepsilon _1}{k_{z}} - i{k_{z1}}{t_1}} \right)}},
		\label{eq:RTM}
	\eeq
\end{widetext}
where
\beq
{k_{z1,2}} = k\sqrt {{\varepsilon _{1,2}} - {{\sin }^2}\theta } , \quad \mbox{Im}\left(k_{z1,2}\right)\ge0,
\label{eq:kz12}
\eeq
are the longitudinal wavenumbers in the two materials and, for notational compactness, we have defined
\beq
{t_{1,2}} = \tan \left( k_{z1,2} d_{1,2} \right),
\label{eq:t12}
\eeq
\beq
{a_{1,2}} = {\varepsilon _1}{k_{z1}}\left( {i{k_{z2}} \pm {\varepsilon _2}{k_{z}}} \right) + \varepsilon _1^2{k_{z}}{k_{z2}}{t_1} \mp i{\varepsilon _2}k_{z1}^2{t_1}.
\eeq
In the ideal case of $\varepsilon_1=0$, it readily follows from Eq. (\ref{eq:RTM}) that the condition $t_2=0$ (corresponding to a Fabry-P\'erot resonance in the core) trivially yields the  coalescence of a zero and a pole, i.e.,  a {\em perfect} BIC, irrespective of cladding thickness $d_1$. This is consistent with results from previous studies.\cite{Sakotic:2020be} For {\em finite} values of the relative permittivities,
by enforcing the zeroing of the numerator in Eq. (\ref{eq:RTM}), we obtain 
\beq
{t_2} = \frac{{2{\varepsilon _1}{\varepsilon _2}{k_{z1}}{k_{z2}}{t_1}\left( {\varepsilon _1^2k_{z}^2 - k_{z1}^2} \right)}}{{\varepsilon _1^2k_{z1}^2\left( {k_{z2}^2 - \varepsilon _2^2k_{z}^2} \right) + t _1^2\left( {\varepsilon _1^4k_{z}^2k_{z2}^2 - \varepsilon _2^2k_{z1}^4} \right)}}.
\label{eq:t2TM}
\eeq
Then, by enforcing the pole condition in Eq. (\ref{eq:RTM}), and substituting the expression in Eq. (\ref{eq:t2TM}), we obtain a quartic equation 
in the unknown $t_1$, which, recalling the branch-cut choice in Eq. (\ref{eq:kz12}), yields the following two classes of solutions
\begin{subequations}
	\begin{eqnarray}
		t_1 &=&   i,
		\quad
		t_2 =  - \frac{2i{\varepsilon _1}{\varepsilon_2}k_{z1}
			k_{z2}}
		{\varepsilon_2^2k_{z1}^2 + \varepsilon_1^2k_{z2}^2},
		\label{eq:TM1}\\
		t_1&=&  - \frac{{i{\varepsilon _1}{k_{z1}}\left( {{\varepsilon _2}{k_{z}} + {k_{z2}}} \right)}}
		{{\varepsilon _1^2{k_{z}}{k_{z2}} + {\varepsilon _2}k_{z1}^2}},\quad
		t_2=  i.
		\label{eq:TM2}
	\end{eqnarray}
\end{subequations}
By inspecting the first class of solutions in Eqs. (\ref{eq:TM1}), and recalling the definitions in Eq. (\ref{eq:t12}), we notice that, if the field is {\em evanescent} in the cladding region (i.e., ${\varepsilon _1} < {\sin ^2}\theta$) and {\em propagating} in the core (i.e., ${\varepsilon_2} > {\sin ^2}\theta$), the second equality can be in principle satisfied for real-valued frequencies. 
Hence, in order to work with arbitrarily small incidence angles, we are led to select an ENZ-type cladding ($\varepsilon_1\ll 1$), whereas the core material can be a regular dielectric.
However, it is apparent that the first equality can never be satisfied exactly (since, for real-valued argument, the hyperbolic tangent is always smaller than one in magnitude), but only {\em asymptotically} (for $d_1\rightarrow\infty$).  This indicates that, for {\em finite} (albeit small) values of the relative permittivities, it is not possible to attain a perfect BIC. Acknowledging this limitation, we focus on suitably approximate solutions (with finite, but arbitrarily large $Q$), which will be referred to as ``quasi-BIC'' (qBIC), which satisfy exactly the zero condition in Eq. (\ref{eq:t2TM}).

To sum up, for given operational frequency, incidence angle, material properties and cladding thickness $d_1$, the relationship in Eq. (\ref{eq:t2TM}) yields the required value of the core thickness $d_2$ to attain a qBIC condition, with the $Q$-factor controlled by $d_1$ and $\varepsilon_1$. This enables a systematic synthesis of qBIC-type responses.

As for the second class of solutions in Eqs. (\ref{eq:TM2}), the role of the core and cladding is reversed. Once again, the second equality can be satisfied only asymptotically (for $d_2\rightarrow\infty$), provided that the field is evanescent in the core region (i.e., ${\varepsilon_2} < {\sin ^2}\theta$). However, unlike the previous case, the right-hand-side in the first equality is generally {\em complex-valued}, which implies that no solutions exist for real-valued frequencies. Therefore, we will not consider this last class of solutions, and focus instead on that in Eqs. (\ref{eq:TM1}).
%
\begin{figure}
	\centering
	\includegraphics[width=\linewidth]{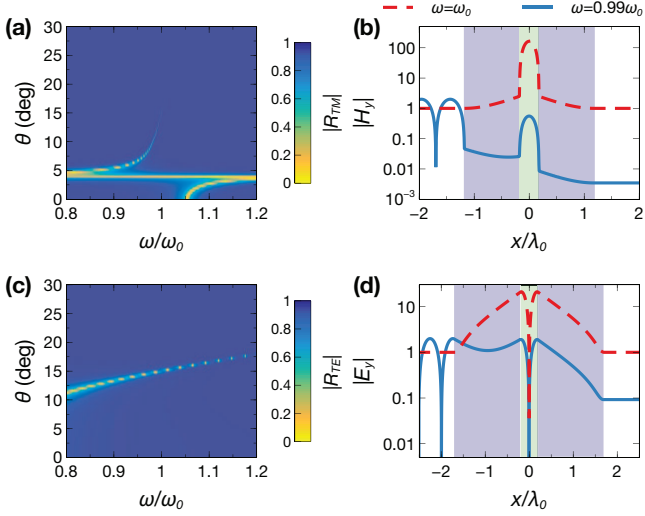}
	\caption{(a) Reflection coefficient magnitude as a function of normalized frequency and incidence angle, for  TM polarization and parameters chosen so as to attain a qBIC at $\omega=\omega_0$ and $\theta=15^o$ ($\varepsilon_1=0.005$, $\varepsilon_2=2$, $d_1=\lambda_0$, $d_2=0.356\lambda_0$). (b) Corresponding field distributions at $\theta=15^o$, with $\omega=\omega_0$ (red-dashed) and $\omega=0.99\omega_0$ (blue-solid); note the semi-log scale. The shaded areas indicate the cladding and core regions. (c), (d) As in panels (a), (b), respectively, but for TE polarization ($d_1=1.5\lambda_0$, $d_2=0.401\lambda_0$).}
	\label{Figure2}
\end{figure}

We now move on to considering the TE polarization, whose interaction with ENZ media is known to be {\em weaker}.\cite{Alu:2007en}
In this case, the reflection coefficient  for the structure in Fig. \ref{Figure1} (for incidence from either side) can be written as
\beq
{R_{^{TE}}}\!=\!\frac{{2{k_{z1}}{k_{z2}}{t_1}\left( {k_{z1}^2\!-\!k_{z}^2} \right)\!+\!{t_2}\left[k_{z1}^2 {\left( {k_{z2}^2\!-\!k_{z}^2} \right) \!+\! t_1^2\left(k_{z}^2k_{z2}^2 \!-\! k_{z1}^4\right)} \right]}}{{t_2}{b_1}{b_2}-{2{k_{z1}}{k_{z2}}\left( {{k_{z1}}{t_1} + i{k_{z}}} \right)\left( {{k_{z1}} - i{k_{z}}{t_1}} \right) }},
\label{eq:RTE}
\eeq
where
\beq
{b_{1,2}} = {k_{z1}}\left( i{{k_{z2}} \pm {k_{z}}} \right)+
t_1
\left( k_z k_{z2} \mp ik_{z1}^2 \right),
\eeq
and all other symbols have already been defined previously. We note that, unlike the TM case, no perfect BIC solutions exist, even in the ideal case $\varepsilon_1=0$. Ideally, in order to enhance the interaction with the TE polarization, and possibly attain a perfect BIC, we should consider (instead, or in addition) a dual, {\em mu-near-zero} condition. 

In the ENZ scenario of interest here, paralleling the reasoning above, we find two classes of solutions
\begin{subequations}
	\begin{eqnarray}
		{t_1} &=&  i,\quad{t_2} =  - \frac{{2i{k_{z1}}{k_{z2}}}}{{k_{z1}^2 + k_{z2}^2}},
		\label{eq:TE1}\\
		{t_1} &=&  \frac{{i{k_{z1}}\left( {{k_{z}} + {k_{z2}}} \right)}}{{k_{z1}^2 + {k_{z}}{k_{z2}}}},\quad{t_2} =i,
		\label{eq:TE2}
	\end{eqnarray}
\end{subequations}
for which qualitatively similar considerations hold as for the TM case above. Specifically, the first class in Eqs.  (\ref{eq:TE1}) leads to a structure with an ENZ cladding. Also in this case, we look for approximate solutions by enforcing the zero condition only, viz.,
\beq
{t_2}=\frac{2{k_{z1}}{k_{z2}}{t_1}\left( {k_{z1}^2\!-\!k_{z}^2} \right)}{k_{z1}^2{\left( {k_{z}^2\!-\!k_{z2}^2} \right) \!+\! t_1^2}\left( {k_{z1}^4 \!-\! k_{z}^2k_{z2}^2} \right)},
\label{eq:t2TE}
\eeq
from which we can derive the core thickness $d_2$, while the cladding thickness $d_1$ controls the $Q$-factor. Even though, as previously mentioned, a perfect BIC is not attainable here, we still refer to this class of solutions as qBIC.
Once again, the second class of solutions in Eqs. (\ref{eq:TE2}) cannot be generally satisfied for real-valued frequencies, and is not pursued here.

Figure \ref{Figure2} shows some representative results for the synthesis of a qBIC at a given angular frequency $\omega_0$ and incidence angle $\theta=15^o$, by assuming $\varepsilon_1=0.005$ and $\varepsilon_2=2$.  Specifically, for the TM polarization, we select $d_1=\lambda_0$ (with $\lambda_0$ denoting the wavelength at the design frequency), and derive from Eq. (\ref{eq:t2TM}) $d_2=0.356\lambda_0$. Figure \ref{Figure2}a shows the corresponding reflection-coefficient magnitude as a function of the normalized frequency and incidence angle, from which the typical signature of a qBIC is visible in terms of a dip with progressively narrower linewidth. For the same parameter configuration, Fig. \ref{Figure2}b shows the spatial field distribution ($|H_y|$) at the resonance, from which we observe a strong field enhancement in the core region, with a typical Fabry-P\'erot pattern, no apparent reflections and full transmission; this is quite different from the response at a nearby frequency (also shown as a reference), from which a strong reflection and very low transmission are instead observed.
Figures \ref{Figure2}c and \ref{Figure2}d show the corresponding results for the TE polarization. Here, assuming $d_1=1.5\lambda_0$, we obtain from Eq. (\ref{eq:t2TE}) $d_2=0.401\lambda_0$. Once again, a qBIC signature is observed, though with a broader linewidth and weaker field enhancement; these characteristics are consistent with what also observed in ENZ slabs.\cite{Alu:2007en}

%
\begin{figure}
	\centering
	\includegraphics[width=\linewidth]{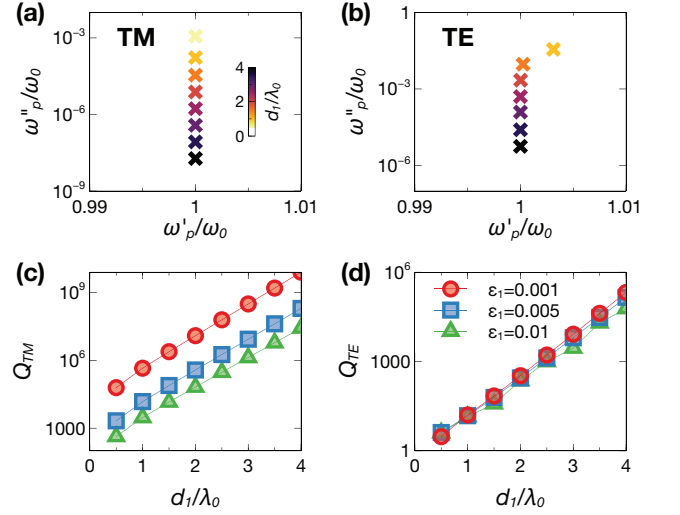}
	\caption{(a), (b) Pole evolution in the complex angular-frequency plane $\tilde\omega_p=\omega^{\prime}_p-i\omega^{\prime\prime}_p$ for TM and TE polarization, respectively, as a function of $d_1/\lambda_0$.
		The cladding and core relative permittivities are fixed at $\varepsilon_1=0.01$ and $\varepsilon_2=2$, respectively, and $d_2$ is computed from Eqs. (\ref{eq:t2TM}) and (\ref{eq:t2TE}), assuming $\theta=15^o$, so as to maintain the zero fixed at $\tilde\omega_p=\omega_0$. (c), (d) Corresponding
		$Q$-factors of qBIC resonances, for three  representative values of $\varepsilon_1$ in the ENZ regime. Note the semi-log scales.}
	\label{Figure3}
\end{figure}

For a more quantitative assessment, we carry out a parametric study of the $Q$-factor, as a function of the thickness and relative permittivity of the cladding region. To this aim, for each combination of $d_1$ and $\varepsilon_1$, we re-calculate $d_2$ from Eqs. (\ref{eq:t2TM}) and (\ref{eq:t2TE}), and numerically compute the poles of the reflection coefficients in Eqs. (\ref{eq:RTM}) and (\ref{eq:RTE}). In view of passivity, this yields a {\em complex-valued} angular resonant frequency $\tilde\omega_p=\omega_p^\prime-i\omega_p^{\prime\prime}$ (with $\omega_p^{\prime\prime}\ge0$), from which we estimate\cite{Christopoulos:2019ot} $Q=\omega_p^\prime/\left(2\omega_p^{\prime\prime}\right)$. Figures \ref{Figure3}a and \ref{Figure3}b show, for both polarizations, a typical pole evolution, with the complex angular frequency asymptotically approaching  $\tilde\omega_p=\omega_0$ (where the zero is placed by design) for increasing values of the the cladding electrical thickness $d_1/\lambda_0$. Figures \ref{Figure3}c and \ref{Figure3}d
show the corresponding $Q$-factors for three representative values of $\varepsilon_1$ in the ENZ regime. In both cases, we observe that the $Q$-factor increases {\em exponentially} with $d_1/\lambda_0$. In particular, for the TM polarization (Fig. \ref{Figure3}c), very high $Q$-factors (>1000) can be attained even for relatively small cladding thicknesses ($d_1\sim0.5\lambda_0$). Moreover, as can be expected, reducing $\varepsilon_1$ (thereby approaching the perfect BIC condition) yields beneficial effects. On the other hand, for the TE polarization (Fig. \ref{Figure3}d), the $Q$-factors are sensibly lower, reaching values >100 for relatively thick claddings ($d_1\gtrsim2\lambda_0$), and the dependence on  $\varepsilon_1$ is much weaker.

%
\begin{figure}
	\centering
	\includegraphics[width=\linewidth]{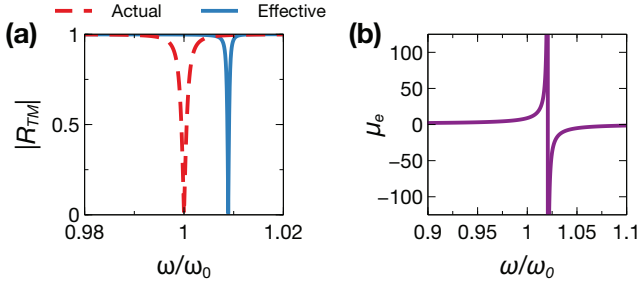}
	\caption{Photonic-doping interpretation. (a) Comparison between the responses (reflection-coefficient magnitude) of the actual (red-dashed) and effective (blue-solid) structures, as a function of normalized frequency. The qBIC resonance is designed so as to occur at $\omega=\omega_0$ and $\theta=5^o$, with parameters: $\varepsilon_1=5\times10^{-4}$, $d_1=0.25\lambda_0$, $\varepsilon_2=2$, $d_2=0.347\lambda_0$. (b) Corresponding effective relative magnetic permeability as a function of normalized frequency.}
	\label{Figure4}
\end{figure}

For the TM polarization, the above results also admit an intriguing interpretation in terms of the ``photonic doping'' concept that was recently put forward by Liberal {\em et al.}\cite{Liberal:2017pd} In essence, they showed that, for a 2-D ENZ host medium and TM polarization, the presence of an arbitrarily placed dielectric particle can dramatically change the optical response, in a manner that resembles the concept of ``doping'' in solid-state physics. Quite interestingly, for an exterior observer, such response is equivalent to an effective magnetic permeability, which can be broadly tuned.
Within this framework, the core region in our configuration (Fig. \ref{Figure1}) can be interpreted as the ``dopant'', and the analytic framework introduced in Ref. \onlinecite{Liberal:2017pd} can be straightforwardly extended to deal with regions of infinite extent. As a result, assuming an ENZ cladding, for an exterior observer, the tri-layer can be effectively replaced by a homogeneous slab of thickness $2d_1+d_2$, same ENZ relative permittivity as the cladding, and effective relative magnetic permeability  
\beq
\mu_{e}=\frac{2\left[k_{z2}d_1+\tan\left(\displaystyle{\frac{k_{z2}d_2}{2}}\right)\right]}{k_{z2}\left(2d_1+d_2\right)}.
\label{eq:mue}
\eeq
Accordingly, the reflection coefficient for the effective structure can be written as
\beq
R_{TM}=\frac{\left(\varepsilon_1^{2} k_z^{2}-k_{ze}^{2}\right) t_{e}}{2 i \varepsilon_1 k_{z} k_{ze}+\left(\varepsilon_1^{2} k_z^{2}+k_{ze}^{2}\right) t_{e}},
\eeq	
where $k_{ze}=k \sqrt{\varepsilon_1 \mu_{e}-\sin^{2} \theta}$, $\mbox{Im}\left(k_{ze}\right)\ge 0$, and 
$t_{e}=\tan\left[k_{ze}\left(2 d_{1}+d_{2}\right)\right]$.
Similar to what observed in Ref. \onlinecite{Liberal:2017pd}, in view of the presence of the tangent term, the effective relative magnetic permeability in Eq. (\ref{eq:mue}) can exhibit {\em arbitrary} values (either positive or negative), and therefore constitutes a key tuning parameter to design a qBIC resonance. Figure \ref{Figure4}a shows a representative comparison between the responses of the actual and effective structures, from which we observe only a very slight frequency shift in the qBIC resonance, attributable to the finite (albeit very small) value of the cladding relative permittivity. Figure \ref{Figure4}b shows instead the corresponding effective relative magnetic permeability, from which we observe  the presence of a pole singularity nearby the qBIC frequency.

%
\begin{figure}
	\centering
	\includegraphics[width=\linewidth]{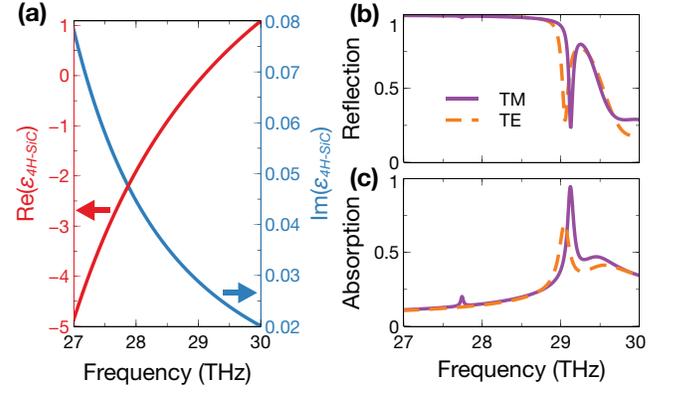}
	\caption{(a) Dispersion law for the 4H-Sic model in Eq. (\ref{eq:SiC}), with $\varepsilon_{\infty}=6.6$, $\gamma=2\pi\cdot0.04$THz, $\omega_{LO}=2\pi\cdot29.08$ THz, and $\omega_{TO}=2\pi\cdot23.89$ THz. Real (red) and imaginary (blue) parts are showed on the left and right axes, respectively. (b), (c) Reflection and absorption coefficient magnitude, respectively, as a function of frequency, for TM (purple-solid) and TE (orange-dashed) polarizations. The qBIC resonances are designed at 29.05 THz and $\theta=15^o$, assuming $\varepsilon_2=2$, $d_1=0.5\lambda_0=5.163\mu m$, and $d_2=0.199\lambda_0=2.05\mu m$ (TM) and $d_2=0.063\lambda_0=0.65\mu m$ (TE).}
	\label{Figure5}
\end{figure}

Finally, we take into account the insofar neglected effects of material dispersion and losses, which are unavoidable, especially in ENZ materials.\cite{Javani:2016ra} As in previous related studies,\cite{Kim:2016ro,Paarman:2016eo} we consider 4H-SiC as a realistic low-loss ENZ material, characterized by a dispersion law
\beq
\varepsilon_{4H-SiC}=\varepsilon_{\infty}\left[1+\frac{\left(\omega_{LO}^{2}-\omega_{TO}^{2}\right)}{\omega_{TO}^{2}-\omega^{2}+i \gamma \omega}\right],
\label{eq:SiC}
\eeq
where $\varepsilon_{\infty}$ denotes the asymptotic (high-frequency) limit, $\gamma$ is a damping coefficient, and $\omega_{LO}$ and $\omega_{TO}$ are the longitudinal and transverse, respectively, optic phonon frequencies. Figure \ref{Figure5}a shows a representative dispersion law within a region of the thermal infrared spectrum, with typical parameters taken from the literature.\cite{Sakotic:2020be} As can be observed, the ENZ condition is reached around 29 THz, with an imaginary part $\mbox{Im}\left(\varepsilon_{4H-SiC}\right)=0.028$.

For these parameters, we carry out the above described syntheses for a qBIC resonance at 29.05 THz and $\theta=15^o$, for both polarizations, assuming $\varepsilon_2=2$ and $d_1=0.5\lambda_0=5.163\mu m$. Clearly, in view of the assumed losses, Eqs. (\ref{eq:t2TM}) and (\ref{eq:t2TE}) now yield {\em complex} values of the core thickness $d_2$, and simply discarding the imaginary part usually leads to poor-quality resonances. However, we found that re-adjusting the real part, by scanning a $\pm 10\%$ neighborhood, is generally sufficient to bring the $Q$-factor to satisfactory levels. Figure \ref{Figure5}b shows two synthesized reflection responses, with $Q$-factors of 509 and 231 for the TM and TE polarization, respectively. As shown in Fig. \ref{Figure5}c, at the designed qBIC angle and frequencies, we also attain sensible and sharp absorption peaks.

To sum up, with focus on symmetric ENZ tri-layers, we have presented a systematic synthesis approach for qBIC-type high-$Q$ resonances at given frequency, incidence angle and polarization. Via parametric studies, we have highlighted the role of the critical parameters and, for the TM polarization, we have provided an insightful phenomenological interpretation in terms of photonic doping. Our results indicate that, even in the presence of realistic lossy materials,  it is possible to attain sharp resonances ($Q$-factor of several hundred) with sensible absorption peaks, for both polarizations. Overall, these outcomes appear of potential interest for applications including light trapping, optical sensing, and narrowband/directive thermal emission.

Among the possible extensions, currently under way, it is worth mentioning the incorporation of gain constituents in order to balance the loss effects, along the lines of previous studies on {\em parity-time} symmetric configurations.\cite{Savoia:2014to,Savoia:2015pt,Sakotic2019:pt,Novitsky:2021qs} Within this framework, also of great interest is the study of cylindrical geometries,\cite{Savoia:2017mi,Moccia:2020hs} also taking into account the recent non-Hermitian extension of the photonic-doping concept.\cite{Coppolaro:2020nh}

\section*{Acknowledgment}
The authors acknowledge partial support from the University of Sannio via the FRA 2019 Program.

\section*{Author Declarations}

\subsection*{Conflict of interest}
The authors have no conflicts to disclose.

\subsection*{Data Availability Statement}
The data that support the findings of this study are available from the corresponding author upon reasonable request.

\appendix

\nocite{*}

\begin{thebibliography}{31}%
	\makeatletter
	\providecommand \@ifxundefined [1]{%
		\@ifx{#1\undefined}
	}%
	\providecommand \@ifnum [1]{%
		\ifnum #1\expandafter \@firstoftwo
		\else \expandafter \@secondoftwo
		\fi
	}%
	\providecommand \@ifx [1]{%
		\ifx #1\expandafter \@firstoftwo
		\else \expandafter \@secondoftwo
		\fi
	}%
	\providecommand \natexlab [1]{#1}%
	\providecommand \enquote  [1]{``#1''}%
	\providecommand \bibnamefont  [1]{#1}%
	\providecommand \bibfnamefont [1]{#1}%
	\providecommand \citenamefont [1]{#1}%
	\providecommand \href@noop [0]{\@secondoftwo}%
	\providecommand \href [0]{\begingroup \@sanitize@url \@href}%
	\providecommand \@href[1]{\@@startlink{#1}\@@href}%
	\providecommand \@@href[1]{\endgroup#1\@@endlink}%
	\providecommand \@sanitize@url [0]{\catcode `\\12\catcode `\$12\catcode
		`\&12\catcode `\#12\catcode `\^12\catcode `\_12\catcode `\%12\relax}%
	\providecommand \@@startlink[1]{}%
	\providecommand \@@endlink[0]{}%
	\providecommand \url  [0]{\begingroup\@sanitize@url \@url }%
	\providecommand \@url [1]{\endgroup\@href {#1}{\urlprefix }}%
	\providecommand \urlprefix  [0]{URL }%
	\providecommand \Eprint [0]{\href }%
	\providecommand \doibase [0]{https://doi.org/}%
	\providecommand \selectlanguage [0]{\@gobble}%
	\providecommand \bibinfo  [0]{\@secondoftwo}%
	\providecommand \bibfield  [0]{\@secondoftwo}%
	\providecommand \translation [1]{[#1]}%
	\providecommand \BibitemOpen [0]{}%
	\providecommand \bibitemStop [0]{}%
	\providecommand \bibitemNoStop [0]{.\EOS\space}%
	\providecommand \EOS [0]{\spacefactor3000\relax}%
	\providecommand \BibitemShut  [1]{\csname bibitem#1\endcsname}%
	\let\auto@bib@innerbib\@empty
	\bibitem [{\citenamefont {{von Neumann}}\ and\ \citenamefont
		{Wigner}(1993)}]{vonNeumann:1993um}%
	\BibitemOpen
	\bibfield  {author} {\bibinfo {author} {\bibfnamefont {J.}~\bibnamefont {{von
					Neumann}}}\ and\ \bibinfo {author} {\bibfnamefont {E.~P.}\ \bibnamefont
			{Wigner}},\ }\bibfield  {title} {\enquote {\bibinfo {title} {\"{U}ber
				{M}erkw\"urdige {D}iskrete {{Eigenwerte}}},}\ }in\ \href@noop {} {\emph
		{\bibinfo {booktitle} {The {{Collected Works}} of {{Eugene Paul Wigner}}}}}\
	(\bibinfo  {publisher} {{Springer}},\ \bibinfo {year} {1993})\ pp.\ \bibinfo
	{pages} {291--293}\BibitemShut {NoStop}%
	\bibitem [{\citenamefont {Marinica}, \citenamefont {Borisov},\ and\
		\citenamefont {Shabanov}(2008)}]{Marinica:2008bs}%
	\BibitemOpen
	\bibfield  {author} {\bibinfo {author} {\bibfnamefont {D.~C.}\ \bibnamefont
			{Marinica}}, \bibinfo {author} {\bibfnamefont {A.~G.}\ \bibnamefont
			{Borisov}},\ and\ \bibinfo {author} {\bibfnamefont {S.~V.}\ \bibnamefont
			{Shabanov}},\ }\bibfield  {title} {\enquote {\bibinfo {title} {Bound states
				in the continuum in photonics},}\ }\href
	{https://doi.org/10.1103/PhysRevLett.100.183902} {\bibfield  {journal}
		{\bibinfo  {journal} {Phys. Rev. Lett.}\ }\textbf {\bibinfo {volume} {100}},\
		\bibinfo {pages} {183902} (\bibinfo {year} {2008})}\BibitemShut {NoStop}%
	\bibitem [{\citenamefont {Plotnik}\ \emph {et~al.}(2011)\citenamefont
		{Plotnik}, \citenamefont {Peleg}, \citenamefont {Dreisow}, \citenamefont
		{Heinrich}, \citenamefont {Nolte}, \citenamefont {Szameit},\ and\
		\citenamefont {Segev}}]{Plotnik:2011eo}%
	\BibitemOpen
	\bibfield  {author} {\bibinfo {author} {\bibfnamefont {Y.}~\bibnamefont
			{Plotnik}}, \bibinfo {author} {\bibfnamefont {O.}~\bibnamefont {Peleg}},
		\bibinfo {author} {\bibfnamefont {F.}~\bibnamefont {Dreisow}}, \bibinfo
		{author} {\bibfnamefont {M.}~\bibnamefont {Heinrich}}, \bibinfo {author}
		{\bibfnamefont {S.}~\bibnamefont {Nolte}}, \bibinfo {author} {\bibfnamefont
			{A.}~\bibnamefont {Szameit}},\ and\ \bibinfo {author} {\bibfnamefont
			{M.}~\bibnamefont {Segev}},\ }\bibfield  {title} {\enquote {\bibinfo {title}
			{Experimental observation of optical bound states in the continuum},}\ }\href
	{https://doi.org/10.1103/PhysRevLett.107.183901} {\bibfield  {journal}
		{\bibinfo  {journal} {Phys. Rev. Lett.}\ }\textbf {\bibinfo {volume} {107}},\
		\bibinfo {pages} {183901} (\bibinfo {year} {2011})}\BibitemShut {NoStop}%
	\bibitem [{\citenamefont {Hsu}\ \emph {et~al.}(2013)\citenamefont {Hsu},
		\citenamefont {Zhen}, \citenamefont {Lee}, \citenamefont {Chua},
		\citenamefont {Johnson}, \citenamefont {Joannopoulos},\ and\ \citenamefont
		{Solja{\v c}i{\'c}}}]{Hsu:2013oo}%
	\BibitemOpen
	\bibfield  {author} {\bibinfo {author} {\bibfnamefont {C.~W.}\ \bibnamefont
			{Hsu}}, \bibinfo {author} {\bibfnamefont {B.}~\bibnamefont {Zhen}}, \bibinfo
		{author} {\bibfnamefont {J.}~\bibnamefont {Lee}}, \bibinfo {author}
		{\bibfnamefont {S.-L.}\ \bibnamefont {Chua}}, \bibinfo {author}
		{\bibfnamefont {S.~G.}\ \bibnamefont {Johnson}}, \bibinfo {author}
		{\bibfnamefont {J.~D.}\ \bibnamefont {Joannopoulos}},\ and\ \bibinfo {author}
		{\bibfnamefont {M.}~\bibnamefont {Solja{\v c}i{\'c}}},\ }\bibfield  {title}
	{\enquote {\bibinfo {title} {Observation of trapped light within the
				radiation continuum},}\ }\href {https://doi.org/10.1038/nature12289}
	{\bibfield  {journal} {\bibinfo  {journal} {Nature}\ }\textbf {\bibinfo
			{volume} {499}},\ \bibinfo {pages} {188--191} (\bibinfo {year}
		{2013})}\BibitemShut {NoStop}%
	\bibitem [{\citenamefont {Silveirinha}(2014)}]{Silveirinha:2014tl}%
	\BibitemOpen
	\bibfield  {author} {\bibinfo {author} {\bibfnamefont {M.~G.}\ \bibnamefont
			{Silveirinha}},\ }\bibfield  {title} {\enquote {\bibinfo {title} {Trapping
				light in open plasmonic nanostructures},}\ }\href
	{https://doi.org/10.1103/PhysRevA.89.023813} {\bibfield  {journal} {\bibinfo
			{journal} {Phys. Rev. A}\ }\textbf {\bibinfo {volume} {89}},\ \bibinfo
		{pages} {023813} (\bibinfo {year} {2014})}\BibitemShut {NoStop}%
	\bibitem [{\citenamefont {Monticone}\ and\ \citenamefont
		{Al\`u}(2014)}]{Monticone:2014ep}%
	\BibitemOpen
	\bibfield  {author} {\bibinfo {author} {\bibfnamefont {F.}~\bibnamefont
			{Monticone}}\ and\ \bibinfo {author} {\bibfnamefont {A.}~\bibnamefont
			{Al\`u}},\ }\bibfield  {title} {\enquote {\bibinfo {title} {Embedded photonic
				eigenvalues in 3{D} nanostructures},}\ }\href
	{https://doi.org/10.1103/PhysRevLett.112.213903} {\bibfield  {journal}
		{\bibinfo  {journal} {Phys. Rev. Lett.}\ }\textbf {\bibinfo {volume} {112}},\
		\bibinfo {pages} {213903} (\bibinfo {year} {2014})}\BibitemShut {NoStop}%
	\bibitem [{\citenamefont {Lanneb{\`e}re}\ and\ \citenamefont
		{Silveirinha}(2015)}]{lannebere:2015om}%
	\BibitemOpen
	\bibfield  {author} {\bibinfo {author} {\bibfnamefont {S.}~\bibnamefont
			{Lanneb{\`e}re}}\ and\ \bibinfo {author} {\bibfnamefont {M.~G.}\ \bibnamefont
			{Silveirinha}},\ }\bibfield  {title} {\enquote {\bibinfo {title} {Optical
				meta-atom for localization of light with quantized energy},}\ }\href
	{https://doi.org/10.1038/ncomms9766} {\bibfield  {journal} {\bibinfo
			{journal} {Nat. Commun.}\ }\textbf {\bibinfo {volume} {6}},\ \bibinfo {pages}
		{8766} (\bibinfo {year} {2015})}\BibitemShut {NoStop}%
	\bibitem [{\citenamefont {Kodigala}\ \emph {et~al.}(2017)\citenamefont
		{Kodigala}, \citenamefont {Lepetit}, \citenamefont {Gu}, \citenamefont
		{Bahari}, \citenamefont {Fainman},\ and\ \citenamefont
		{Kant{\'e}}}]{Kodigala:2017la}%
	\BibitemOpen
	\bibfield  {author} {\bibinfo {author} {\bibfnamefont {A.}~\bibnamefont
			{Kodigala}}, \bibinfo {author} {\bibfnamefont {T.}~\bibnamefont {Lepetit}},
		\bibinfo {author} {\bibfnamefont {Q.}~\bibnamefont {Gu}}, \bibinfo {author}
		{\bibfnamefont {B.}~\bibnamefont {Bahari}}, \bibinfo {author} {\bibfnamefont
			{Y.}~\bibnamefont {Fainman}},\ and\ \bibinfo {author} {\bibfnamefont
			{B.}~\bibnamefont {Kant{\'e}}},\ }\bibfield  {title} {\enquote {\bibinfo
			{title} {Lasing action from photonic bound states in continuum},}\ }\href
	{https://doi.org/10.1038/nature20799} {\bibfield  {journal} {\bibinfo
			{journal} {Nature}\ }\textbf {\bibinfo {volume} {541}},\ \bibinfo {pages}
		{196--199} (\bibinfo {year} {2017})}\BibitemShut {NoStop}%
	\bibitem [{\citenamefont {Rybin}\ \emph {et~al.}(2017)\citenamefont {Rybin},
		\citenamefont {Koshelev}, \citenamefont {Sadrieva}, \citenamefont {Samusev},
		\citenamefont {Bogdanov}, \citenamefont {Limonov},\ and\ \citenamefont
		{Kivshar}}]{Rybin:2017hq}%
	\BibitemOpen
	\bibfield  {author} {\bibinfo {author} {\bibfnamefont {M.~V.}\ \bibnamefont
			{Rybin}}, \bibinfo {author} {\bibfnamefont {K.~L.}\ \bibnamefont {Koshelev}},
		\bibinfo {author} {\bibfnamefont {Z.~F.}\ \bibnamefont {Sadrieva}}, \bibinfo
		{author} {\bibfnamefont {K.~B.}\ \bibnamefont {Samusev}}, \bibinfo {author}
		{\bibfnamefont {A.~A.}\ \bibnamefont {Bogdanov}}, \bibinfo {author}
		{\bibfnamefont {M.~F.}\ \bibnamefont {Limonov}},\ and\ \bibinfo {author}
		{\bibfnamefont {Y.~S.}\ \bibnamefont {Kivshar}},\ }\bibfield  {title}
	{\enquote {\bibinfo {title} {High-${Q}$ supercavity modes in subwavelength
				dielectric resonators},}\ }\href
	{https://doi.org/10.1103/PhysRevLett.119.243901} {\bibfield  {journal}
		{\bibinfo  {journal} {Phys. Rev. Lett.}\ }\textbf {\bibinfo {volume} {119}},\
		\bibinfo {pages} {243901} (\bibinfo {year} {2017})}\BibitemShut {NoStop}%
	\bibitem [{\citenamefont {Bezus}, \citenamefont {Bykov},\ and\ \citenamefont
		{Doskolovich}(2018)}]{Bezus:2018bs}%
	\BibitemOpen
	\bibfield  {author} {\bibinfo {author} {\bibfnamefont {E.~A.}\ \bibnamefont
			{Bezus}}, \bibinfo {author} {\bibfnamefont {D.~A.}\ \bibnamefont {Bykov}},\
		and\ \bibinfo {author} {\bibfnamefont {L.~L.}\ \bibnamefont {Doskolovich}},\
	}\bibfield  {title} {\enquote {\bibinfo {title} {Bound states in the
				continuum and high-q resonances supported by a dielectric ridge on a slab
				waveguide},}\ }\href {https://doi.org/10.1364/PRJ.6.001084} {\bibfield
		{journal} {\bibinfo  {journal} {Photon. Res.}\ }\textbf {\bibinfo {volume}
			{6}},\ \bibinfo {pages} {1084--1093} (\bibinfo {year} {2018})}\BibitemShut
	{NoStop}%
	\bibitem [{\citenamefont {Koshelev}\ \emph {et~al.}(2018)\citenamefont
		{Koshelev}, \citenamefont {Lepeshov}, \citenamefont {Liu}, \citenamefont
		{Bogdanov},\ and\ \citenamefont {Kivshar}}]{Koshelev:2018am}%
	\BibitemOpen
	\bibfield  {author} {\bibinfo {author} {\bibfnamefont {K.}~\bibnamefont
			{Koshelev}}, \bibinfo {author} {\bibfnamefont {S.}~\bibnamefont {Lepeshov}},
		\bibinfo {author} {\bibfnamefont {M.}~\bibnamefont {Liu}}, \bibinfo {author}
		{\bibfnamefont {A.}~\bibnamefont {Bogdanov}},\ and\ \bibinfo {author}
		{\bibfnamefont {Y.}~\bibnamefont {Kivshar}},\ }\bibfield  {title} {\enquote
		{\bibinfo {title} {Asymmetric metasurfaces with high-${Q}$ resonances
				governed by bound states in the continuum},}\ }\href
	{https://doi.org/10.1103/PhysRevLett.121.193903} {\bibfield  {journal}
		{\bibinfo  {journal} {Phys. Rev. Lett.}\ }\textbf {\bibinfo {volume} {121}},\
		\bibinfo {pages} {193903} (\bibinfo {year} {2018})}\BibitemShut {NoStop}%
	\bibitem [{\citenamefont {Monticone}\ \emph {et~al.}(2018)\citenamefont
		{Monticone}, \citenamefont {Doeleman}, \citenamefont {Den~Hollander},
		\citenamefont {Koenderink},\ and\ \citenamefont
		{Al\`{u}}}]{Monticone:2018tl}%
	\BibitemOpen
	\bibfield  {author} {\bibinfo {author} {\bibfnamefont {F.}~\bibnamefont
			{Monticone}}, \bibinfo {author} {\bibfnamefont {H.~M.}\ \bibnamefont
			{Doeleman}}, \bibinfo {author} {\bibfnamefont {W.}~\bibnamefont
			{Den~Hollander}}, \bibinfo {author} {\bibfnamefont {A.~F.}\ \bibnamefont
			{Koenderink}},\ and\ \bibinfo {author} {\bibfnamefont {A.}~\bibnamefont
			{Al\`{u}}},\ }\bibfield  {title} {\enquote {\bibinfo {title} {Trapping light
				in plain sight: Embedded photonic eigenstates in zero-index metamaterials},}\
	}\href {https://doi.org/https://doi.org/10.1002/lpor.201700220} {\bibfield
		{journal} {\bibinfo  {journal} {Laser \& Photon. Rev.}\ }\textbf {\bibinfo
			{volume} {12}},\ \bibinfo {pages} {1700220} (\bibinfo {year}
		{2018})}\BibitemShut {NoStop}%
	\bibitem [{\citenamefont {Sakotic}\ \emph {et~al.}(2020)\citenamefont
		{Sakotic}, \citenamefont {Krasnok}, \citenamefont {Cselyuszka}, \citenamefont
		{Jankovic},\ and\ \citenamefont {Al\`u}}]{Sakotic:2020be}%
	\BibitemOpen
	\bibfield  {author} {\bibinfo {author} {\bibfnamefont {Z.}~\bibnamefont
			{Sakotic}}, \bibinfo {author} {\bibfnamefont {A.}~\bibnamefont {Krasnok}},
		\bibinfo {author} {\bibfnamefont {N.}~\bibnamefont {Cselyuszka}}, \bibinfo
		{author} {\bibfnamefont {N.}~\bibnamefont {Jankovic}},\ and\ \bibinfo
		{author} {\bibfnamefont {A.}~\bibnamefont {Al\`u}},\ }\bibfield  {title}
	{\enquote {\bibinfo {title} {Berreman embedded eigenstates for narrow-band
				absorption and thermal emission},}\ }\href
	{https://doi.org/10.1103/PhysRevApplied.13.064073} {\bibfield  {journal}
		{\bibinfo  {journal} {Phys. Rev. Appl.}\ }\textbf {\bibinfo {volume} {13}},\
		\bibinfo {pages} {064073} (\bibinfo {year} {2020})}\BibitemShut {NoStop}%
	\bibitem [{\citenamefont {Hsu}\ \emph {et~al.}(2016)\citenamefont {Hsu},
		\citenamefont {Zhen}, \citenamefont {Stone}, \citenamefont {Joannopoulos},\
		and\ \citenamefont {Solja{\v c}i{\'c}}}]{Hsu:2016bs}%
	\BibitemOpen
	\bibfield  {author} {\bibinfo {author} {\bibfnamefont {C.~W.}\ \bibnamefont
			{Hsu}}, \bibinfo {author} {\bibfnamefont {B.}~\bibnamefont {Zhen}}, \bibinfo
		{author} {\bibfnamefont {A.~D.}\ \bibnamefont {Stone}}, \bibinfo {author}
		{\bibfnamefont {J.~D.}\ \bibnamefont {Joannopoulos}},\ and\ \bibinfo {author}
		{\bibfnamefont {M.}~\bibnamefont {Solja{\v c}i{\'c}}},\ }\bibfield  {title}
	{\enquote {\bibinfo {title} {Bound states in the continuum},}\ }\href
	{https://doi.org/10.1038/natrevmats.2016.48} {\bibfield  {journal} {\bibinfo
			{journal} {Nat. Rev. Mater.}\ }\textbf {\bibinfo {volume} {1}},\ \bibinfo
		{pages} {16048} (\bibinfo {year} {2016})}\BibitemShut {NoStop}%
	\bibitem [{\citenamefont {Krasnok}\ \emph {et~al.}(2019)\citenamefont
		{Krasnok}, \citenamefont {Baranov}, \citenamefont {Li}, \citenamefont {Miri},
		\citenamefont {Monticone},\ and\ \citenamefont {Al\`{u}}}]{Krasnok:2019ai}%
	\BibitemOpen
	\bibfield  {author} {\bibinfo {author} {\bibfnamefont {A.}~\bibnamefont
			{Krasnok}}, \bibinfo {author} {\bibfnamefont {D.}~\bibnamefont {Baranov}},
		\bibinfo {author} {\bibfnamefont {H.}~\bibnamefont {Li}}, \bibinfo {author}
		{\bibfnamefont {M.-A.}\ \bibnamefont {Miri}}, \bibinfo {author}
		{\bibfnamefont {F.}~\bibnamefont {Monticone}},\ and\ \bibinfo {author}
		{\bibfnamefont {A.}~\bibnamefont {Al\`{u}}},\ }\bibfield  {title} {\enquote
		{\bibinfo {title} {Anomalies in light scattering},}\ }\href
	{https://doi.org/10.1364/AOP.11.000892} {\bibfield  {journal} {\bibinfo
			{journal} {Adv. Opt. Photon.}\ }\textbf {\bibinfo {volume} {11}},\ \bibinfo
		{pages} {892--951} (\bibinfo {year} {2019})}\BibitemShut {NoStop}%
	\bibitem [{\citenamefont {Azzam}\ and\ \citenamefont
		{Kildishev}(2021)}]{Azzam:2021pb}%
	\BibitemOpen
	\bibfield  {author} {\bibinfo {author} {\bibfnamefont {S.~I.}\ \bibnamefont
			{Azzam}}\ and\ \bibinfo {author} {\bibfnamefont {A.~V.}\ \bibnamefont
			{Kildishev}},\ }\bibfield  {title} {\enquote {\bibinfo {title} {Photonic
				bound states in the continuum: From basics to applications},}\ }\href
	{https://doi.org/10.1002/adom.202001469} {\bibfield  {journal} {\bibinfo
			{journal} {Adv. Opt. Mater.}\ }\textbf {\bibinfo {volume} {9}},\ \bibinfo
		{pages} {2001469} (\bibinfo {year} {2021})}\BibitemShut {NoStop}%
	\bibitem [{\citenamefont {Liberal}\ and\ \citenamefont
		{Engheta}(2017)}]{Liberal:2017nz}%
	\BibitemOpen
	\bibfield  {author} {\bibinfo {author} {\bibfnamefont {I.}~\bibnamefont
			{Liberal}}\ and\ \bibinfo {author} {\bibfnamefont {N.}~\bibnamefont
			{Engheta}},\ }\bibfield  {title} {\enquote {\bibinfo {title} {Near-zero
				refractive index photonics},}\ }\href
	{https://doi.org/10.1038/nphoton.2017.13} {\bibfield  {journal} {\bibinfo
			{journal} {Nat. Photon.}\ }\textbf {\bibinfo {volume} {11}},\ \bibinfo
		{pages} {149--158} (\bibinfo {year} {2017})}\BibitemShut {NoStop}%
	\bibitem [{\citenamefont {Wu}\ \emph {et~al.}(2021)\citenamefont {Wu},
		\citenamefont {Xie}, \citenamefont {Sha}, \citenamefont {Fu},\ and\
		\citenamefont {Li}}]{Wu:2021en}%
	\BibitemOpen
	\bibfield  {author} {\bibinfo {author} {\bibfnamefont {J.}~\bibnamefont
			{Wu}}, \bibinfo {author} {\bibfnamefont {Z.~T.}\ \bibnamefont {Xie}},
		\bibinfo {author} {\bibfnamefont {Y.}~\bibnamefont {Sha}}, \bibinfo {author}
		{\bibfnamefont {H.~Y.}\ \bibnamefont {Fu}},\ and\ \bibinfo {author}
		{\bibfnamefont {Q.}~\bibnamefont {Li}},\ }\bibfield  {title} {\enquote
		{\bibinfo {title} {Epsilon-near-zero photonics: infinite potentials},}\
	}\href {https://doi.org/10.1364/PRJ.427246} {\bibfield  {journal} {\bibinfo
			{journal} {Photon. Res.}\ }\textbf {\bibinfo {volume} {9}},\ \bibinfo {pages}
		{1616--1644} (\bibinfo {year} {2021})}\BibitemShut {NoStop}%
	\bibitem [{\citenamefont {Liberal}\ \emph {et~al.}(2017)\citenamefont
		{Liberal}, \citenamefont {Mahmoud}, \citenamefont {Li}, \citenamefont
		{Edwards},\ and\ \citenamefont {Engheta}}]{Liberal:2017pd}%
	\BibitemOpen
	\bibfield  {author} {\bibinfo {author} {\bibfnamefont {I.}~\bibnamefont
			{Liberal}}, \bibinfo {author} {\bibfnamefont {A.~M.}\ \bibnamefont
			{Mahmoud}}, \bibinfo {author} {\bibfnamefont {Y.}~\bibnamefont {Li}},
		\bibinfo {author} {\bibfnamefont {B.}~\bibnamefont {Edwards}},\ and\ \bibinfo
		{author} {\bibfnamefont {N.}~\bibnamefont {Engheta}},\ }\bibfield  {title}
	{\enquote {\bibinfo {title} {Photonic doping of epsilon-near-zero media},}\
	}\href {https://doi.org/10.1126/science.aal2672} {\bibfield  {journal}
		{\bibinfo  {journal} {Science}\ }\textbf {\bibinfo {volume} {355}},\ \bibinfo
		{pages} {1058--1062} (\bibinfo {year} {2017})}\BibitemShut {NoStop}%
	\bibitem [{\citenamefont {Savoia}\ \emph {et~al.}(2014)\citenamefont {Savoia},
		\citenamefont {Castaldi}, \citenamefont {Galdi}, \citenamefont {Al\`u},\ and\
		\citenamefont {Engheta}}]{Savoia:2014to}%
	\BibitemOpen
	\bibfield  {author} {\bibinfo {author} {\bibfnamefont {S.}~\bibnamefont
			{Savoia}}, \bibinfo {author} {\bibfnamefont {G.}~\bibnamefont {Castaldi}},
		\bibinfo {author} {\bibfnamefont {V.}~\bibnamefont {Galdi}}, \bibinfo
		{author} {\bibfnamefont {A.}~\bibnamefont {Al\`u}},\ and\ \bibinfo {author}
		{\bibfnamefont {N.}~\bibnamefont {Engheta}},\ }\bibfield  {title} {\enquote
		{\bibinfo {title} {Tunneling of obliquely incident waves through
				$\mathcal{PT}$-symmetric epsilon-near-zero bilayers},}\ }\href
	{https://doi.org/10.1103/PhysRevB.89.085105} {\bibfield  {journal} {\bibinfo
			{journal} {Phys. Rev. B}\ }\textbf {\bibinfo {volume} {89}},\ \bibinfo
		{pages} {085105} (\bibinfo {year} {2014})}\BibitemShut {NoStop}%
	\bibitem [{\citenamefont {Al\`u}\ \emph {et~al.}(2007)\citenamefont {Al\`u},
		\citenamefont {Silveirinha}, \citenamefont {Salandrino},\ and\ \citenamefont
		{Engheta}}]{Alu:2007en}%
	\BibitemOpen
	\bibfield  {author} {\bibinfo {author} {\bibfnamefont {A.}~\bibnamefont
			{Al\`u}}, \bibinfo {author} {\bibfnamefont {M.~G.}\ \bibnamefont
			{Silveirinha}}, \bibinfo {author} {\bibfnamefont {A.}~\bibnamefont
			{Salandrino}},\ and\ \bibinfo {author} {\bibfnamefont {N.}~\bibnamefont
			{Engheta}},\ }\bibfield  {title} {\enquote {\bibinfo {title}
			{Epsilon-near-zero metamaterials and electromagnetic sources: Tailoring the
				radiation phase pattern},}\ }\href
	{https://doi.org/10.1103/PhysRevB.75.155410} {\bibfield  {journal} {\bibinfo
			{journal} {Phys. Rev. B}\ }\textbf {\bibinfo {volume} {75}},\ \bibinfo
		{pages} {155410} (\bibinfo {year} {2007})}\BibitemShut {NoStop}%
	\bibitem [{\citenamefont {Christopoulos}\ \emph {et~al.}(2019)\citenamefont
		{Christopoulos}, \citenamefont {Tsilipakos}, \citenamefont {Sinatkas},\ and\
		\citenamefont {Kriezis}}]{Christopoulos:2019ot}%
	\BibitemOpen
	\bibfield  {author} {\bibinfo {author} {\bibfnamefont {T.}~\bibnamefont
			{Christopoulos}}, \bibinfo {author} {\bibfnamefont {O.}~\bibnamefont
			{Tsilipakos}}, \bibinfo {author} {\bibfnamefont {G.}~\bibnamefont
			{Sinatkas}},\ and\ \bibinfo {author} {\bibfnamefont {E.~E.}\ \bibnamefont
			{Kriezis}},\ }\bibfield  {title} {\enquote {\bibinfo {title} {On the
				calculation of the quality factor in contemporary photonic resonant
				structures},}\ }\href {https://doi.org/10.1364/OE.27.014505} {\bibfield
		{journal} {\bibinfo  {journal} {Opt. Express}\ }\textbf {\bibinfo {volume}
			{27}},\ \bibinfo {pages} {14505--14522} (\bibinfo {year} {2019})}\BibitemShut
	{NoStop}%
	\bibitem [{\citenamefont {Javani}\ and\ \citenamefont
		{Stockman}(2016)}]{Javani:2016ra}%
	\BibitemOpen
	\bibfield  {author} {\bibinfo {author} {\bibfnamefont {M.~H.}\ \bibnamefont
			{Javani}}\ and\ \bibinfo {author} {\bibfnamefont {M.~I.}\ \bibnamefont
			{Stockman}},\ }\bibfield  {title} {\enquote {\bibinfo {title} {Real and
				imaginary properties of epsilon-near-zero materials},}\ }\href
	{https://doi.org/10.1103/PhysRevLett.117.107404} {\bibfield  {journal}
		{\bibinfo  {journal} {Phys. Rev. Lett.}\ }\textbf {\bibinfo {volume} {117}},\
		\bibinfo {pages} {107404} (\bibinfo {year} {2016})}\BibitemShut {NoStop}%
	\bibitem [{\citenamefont {Kim}\ \emph {et~al.}(2016)\citenamefont {Kim},
		\citenamefont {Dutta}, \citenamefont {Naik}, \citenamefont {Giles},
		\citenamefont {Bezares}, \citenamefont {Ellis}, \citenamefont {Tischler},
		\citenamefont {Mahmoud}, \citenamefont {Caglayan}, \citenamefont {Glembocki},
		\citenamefont {Kildishev}, \citenamefont {Caldwell}, \citenamefont
		{Boltasseva},\ and\ \citenamefont {Engheta}}]{Kim:2016ro}%
	\BibitemOpen
	\bibfield  {author} {\bibinfo {author} {\bibfnamefont {J.}~\bibnamefont
			{Kim}}, \bibinfo {author} {\bibfnamefont {A.}~\bibnamefont {Dutta}}, \bibinfo
		{author} {\bibfnamefont {G.~V.}\ \bibnamefont {Naik}}, \bibinfo {author}
		{\bibfnamefont {A.~J.}\ \bibnamefont {Giles}}, \bibinfo {author}
		{\bibfnamefont {F.~J.}\ \bibnamefont {Bezares}}, \bibinfo {author}
		{\bibfnamefont {C.~T.}\ \bibnamefont {Ellis}}, \bibinfo {author}
		{\bibfnamefont {J.~G.}\ \bibnamefont {Tischler}}, \bibinfo {author}
		{\bibfnamefont {A.~M.}\ \bibnamefont {Mahmoud}}, \bibinfo {author}
		{\bibfnamefont {H.}~\bibnamefont {Caglayan}}, \bibinfo {author}
		{\bibfnamefont {O.~J.}\ \bibnamefont {Glembocki}}, \bibinfo {author}
		{\bibfnamefont {A.~V.}\ \bibnamefont {Kildishev}}, \bibinfo {author}
		{\bibfnamefont {J.~D.}\ \bibnamefont {Caldwell}}, \bibinfo {author}
		{\bibfnamefont {A.}~\bibnamefont {Boltasseva}},\ and\ \bibinfo {author}
		{\bibfnamefont {N.}~\bibnamefont {Engheta}},\ }\bibfield  {title} {\enquote
		{\bibinfo {title} {Role of epsilon-near-zero substrates in the optical
				response of plasmonic antennas},}\ }\href
	{https://doi.org/10.1364/OPTICA.3.000339} {\bibfield  {journal} {\bibinfo
			{journal} {Optica}\ }\textbf {\bibinfo {volume} {3}},\ \bibinfo {pages}
		{339--346} (\bibinfo {year} {2016})}\BibitemShut {NoStop}%
	\bibitem [{\citenamefont {Paarmann}\ \emph {et~al.}(2016)\citenamefont
		{Paarmann}, \citenamefont {Razdolski}, \citenamefont {Gewinner},
		\citenamefont {Sch\"ollkopf},\ and\ \citenamefont {Wolf}}]{Paarman:2016eo}%
	\BibitemOpen
	\bibfield  {author} {\bibinfo {author} {\bibfnamefont {A.}~\bibnamefont
			{Paarmann}}, \bibinfo {author} {\bibfnamefont {I.}~\bibnamefont {Razdolski}},
		\bibinfo {author} {\bibfnamefont {S.}~\bibnamefont {Gewinner}}, \bibinfo
		{author} {\bibfnamefont {W.}~\bibnamefont {Sch\"ollkopf}},\ and\ \bibinfo
		{author} {\bibfnamefont {M.}~\bibnamefont {Wolf}},\ }\bibfield  {title}
	{\enquote {\bibinfo {title} {Effects of crystal anisotropy on optical phonon
				resonances in midinfrared second harmonic response of {SiC}},}\ }\href
	{https://doi.org/10.1103/PhysRevB.94.134312} {\bibfield  {journal} {\bibinfo
			{journal} {Phys. Rev. B}\ }\textbf {\bibinfo {volume} {94}},\ \bibinfo
		{pages} {134312} (\bibinfo {year} {2016})}\BibitemShut {NoStop}%
	\bibitem [{\citenamefont {Savoia}\ \emph {et~al.}(2015)\citenamefont {Savoia},
		\citenamefont {Castaldi}, \citenamefont {Galdi}, \citenamefont {Al\`u},\ and\
		\citenamefont {Engheta}}]{Savoia:2015pt}%
	\BibitemOpen
	\bibfield  {author} {\bibinfo {author} {\bibfnamefont {S.}~\bibnamefont
			{Savoia}}, \bibinfo {author} {\bibfnamefont {G.}~\bibnamefont {Castaldi}},
		\bibinfo {author} {\bibfnamefont {V.}~\bibnamefont {Galdi}}, \bibinfo
		{author} {\bibfnamefont {A.}~\bibnamefont {Al\`u}},\ and\ \bibinfo {author}
		{\bibfnamefont {N.}~\bibnamefont {Engheta}},\ }\bibfield  {title} {\enquote
		{\bibinfo {title} {$\mathcal{PT}$-symmetry-induced wave confinement and
				guiding in $\ensuremath{\epsilon}$-near-zero metamaterials},}\ }\href
	{https://doi.org/10.1103/PhysRevB.91.115114} {\bibfield  {journal} {\bibinfo
			{journal} {Phys. Rev. B}\ }\textbf {\bibinfo {volume} {91}},\ \bibinfo
		{pages} {115114} (\bibinfo {year} {2015})}\BibitemShut {NoStop}%
	\bibitem [{\citenamefont {Sakotic}\ \emph {et~al.}(2019)\citenamefont
		{Sakotic}, \citenamefont {Krasnok}, \citenamefont {Cselyuszka}, \citenamefont
		{Jankovic},\ and\ \citenamefont {Al{\`u}}}]{Sakotic2019:pt}%
	\BibitemOpen
	\bibfield  {author} {\bibinfo {author} {\bibfnamefont {Z.}~\bibnamefont
			{Sakotic}}, \bibinfo {author} {\bibfnamefont {A.}~\bibnamefont {Krasnok}},
		\bibinfo {author} {\bibfnamefont {N.}~\bibnamefont {Cselyuszka}}, \bibinfo
		{author} {\bibfnamefont {N.}~\bibnamefont {Jankovic}},\ and\ \bibinfo
		{author} {\bibfnamefont {A.}~\bibnamefont {Al{\`u}}},\ }\bibfield  {title}
	{\enquote {\bibinfo {title} {{PT}-symmetric cladding layers for high-{Q}
				{B}rewster modes and embedded eigenstates},}\ }in\ \href
	{https://doi.org/10.1109/MetaMaterials.2019.8900850} {\emph {\bibinfo
			{booktitle} {Proc. Metamaterials}}}\ (\bibinfo {year} {2019})\ pp.\
	\bibinfo {pages} {X--357--X--359}\BibitemShut {NoStop}%
	\bibitem [{\citenamefont {Novitsky}\ \emph {et~al.}(2021)\citenamefont
		{Novitsky}, \citenamefont {Shalin}, \citenamefont {Redka}, \citenamefont
		{Bobrovs},\ and\ \citenamefont {Novitsky}}]{Novitsky:2021qs}%
	\BibitemOpen
	\bibfield  {author} {\bibinfo {author} {\bibfnamefont {D.~V.}\ \bibnamefont
			{Novitsky}}, \bibinfo {author} {\bibfnamefont {A.~S.}\ \bibnamefont
			{Shalin}}, \bibinfo {author} {\bibfnamefont {D.}~\bibnamefont {Redka}},
		\bibinfo {author} {\bibfnamefont {V.}~\bibnamefont {Bobrovs}},\ and\ \bibinfo
		{author} {\bibfnamefont {A.~V.}\ \bibnamefont {Novitsky}},\ }\bibfield
	{title} {\enquote {\bibinfo {title} {Quasibound states in the continuum
				induced by $\mathcal{PT}$ symmetry breaking},}\ }\href
	{https://doi.org/10.1103/PhysRevB.104.085126} {\bibfield  {journal} {\bibinfo
			{journal} {Phys. Rev. B}\ }\textbf {\bibinfo {volume} {104}},\ \bibinfo
		{pages} {085126} (\bibinfo {year} {2021})}\BibitemShut {NoStop}%
	\bibitem [{\citenamefont {Savoia}\ \emph {et~al.}(2017)\citenamefont {Savoia},
		\citenamefont {Valagiannopoulos}, \citenamefont {Monticone}, \citenamefont
		{Castaldi}, \citenamefont {Galdi},\ and\ \citenamefont
		{Al{\`u}}}]{Savoia:2017mi}%
	\BibitemOpen
	\bibfield  {author} {\bibinfo {author} {\bibfnamefont {S.}~\bibnamefont
			{Savoia}}, \bibinfo {author} {\bibfnamefont {C.~A.}\ \bibnamefont
			{Valagiannopoulos}}, \bibinfo {author} {\bibfnamefont {F.}~\bibnamefont
			{Monticone}}, \bibinfo {author} {\bibfnamefont {G.}~\bibnamefont {Castaldi}},
		\bibinfo {author} {\bibfnamefont {V.}~\bibnamefont {Galdi}},\ and\ \bibinfo
		{author} {\bibfnamefont {A.}~\bibnamefont {Al{\`u}}},\ }\bibfield  {title}
	{\enquote {\bibinfo {title} {Magnified imaging based on non-{H}ermitian
				nonlocal cylindrical metasurfaces},}\ }\href
	{https://doi.org/10.1103/PhysRevB.95.115114} {\bibfield  {journal} {\bibinfo
			{journal} {Phys. Rev. B}\ }\textbf {\bibinfo {volume} {95}},\ \bibinfo
		{pages} {115114--13} (\bibinfo {year} {2017})}\BibitemShut {NoStop}%
	\bibitem [{\citenamefont {Moccia}\ \emph {et~al.}(2020)\citenamefont {Moccia},
		\citenamefont {Castaldi}, \citenamefont {Al\`u},\ and\ \citenamefont
		{Galdi}}]{Moccia:2020hs}%
	\BibitemOpen
	\bibfield  {author} {\bibinfo {author} {\bibfnamefont {M.}~\bibnamefont
			{Moccia}}, \bibinfo {author} {\bibfnamefont {G.}~\bibnamefont {Castaldi}},
		\bibinfo {author} {\bibfnamefont {A.}~\bibnamefont {Al\`u}},\ and\ \bibinfo
		{author} {\bibfnamefont {V.}~\bibnamefont {Galdi}},\ }\bibfield  {title}
	{\enquote {\bibinfo {title} {Harnessing spectral singularities in
				non-{H}ermitian cylindrical structures},}\ }\href
	{https://doi.org/10.1109/TAP.2019.2927861} {\bibfield  {journal} {\bibinfo
			{journal} {IEEE Trans. Antennas Propagat.}\ }\textbf {\bibinfo {volume}
			{68}},\ \bibinfo {pages} {1704--1716} (\bibinfo {year} {2020})}\BibitemShut
	{NoStop}%
	\bibitem [{\citenamefont {Coppolaro}\ \emph {et~al.}(2020)\citenamefont
		{Coppolaro}, \citenamefont {Moccia}, \citenamefont {Castaldi}, \citenamefont
		{Engheta},\ and\ \citenamefont {Galdi}}]{Coppolaro:2020nh}%
	\BibitemOpen
	\bibfield  {author} {\bibinfo {author} {\bibfnamefont {M.}~\bibnamefont
			{Coppolaro}}, \bibinfo {author} {\bibfnamefont {M.}~\bibnamefont {Moccia}},
		\bibinfo {author} {\bibfnamefont {G.}~\bibnamefont {Castaldi}}, \bibinfo
		{author} {\bibfnamefont {N.}~\bibnamefont {Engheta}},\ and\ \bibinfo {author}
		{\bibfnamefont {V.}~\bibnamefont {Galdi}},\ }\bibfield  {title} {\enquote
		{\bibinfo {title} {Non-{H}ermitian doping of epsilon-near-zero media},}\
	}\href {https://doi.org/10.1073/pnas.2001125117} {\bibfield  {journal}
		{\bibinfo  {journal} {Proc. Natl. Acad. Sci. U.S.A.}\ }\textbf {\bibinfo
			{volume} {117}},\ \bibinfo {pages} {13921--13928} (\bibinfo {year}
		{2020})}\BibitemShut {NoStop}%
\end{thebibliography}

%

\end{document}